# Learning and Spatiotemporally Correlated Functions Mimicked in Oxide-Based Artificial Synaptic Transistors

Chang Jin Wan[1,2], Li Qiang Zhu[2], Yi Shi[1], and Qing Wan[1,2,*]

1) School of Electronic Science & Engineering, Nanjing University, Nanjing 210093, Peoples Republic of China

2) Ningbo Institute of Materials Technology and Engineering, Chinese Academy of Sciences, Ningbo 315201, People's Republic of China

**Abstract**

Learning and logic are fundamental brain functions that make the individual to adapt to the environment, and such functions are established in human brain by modulating ionic fluxes in synapses. Nanoscale ionic/electronic devices with inherent synaptic functions are considered to be essential building blocks for artificial neural networks. Here, Multi-terminal IZO-based artificial synaptic transistors gated by fast proton-conducting phosphosilicate electrolytes are fabricated on glass substrates. Proton in the $SiO_2$ electrolyte and IZO channel conductance are regarded as the neurotransmitter and synaptic weight, respectively. Spike-timing dependent plasticity, short-term memory and long-term memory were successfully mimicked in such protonic/electronic hybrid artificial synapses. And most importantly, spatiotemporally correlated logic functions are also mimicked in a simple artificial neural network without any intentional hard-wire connections due to the naturally proton-related coupling effect. The oxide-based protonic/electronic hybrid artificial synaptic transistors reported here are potential building blocks for artificial neural networks.

---

Learning and logic are fundamental brain functions that make the individual to adapt to the environment, and such functions are established in human brain by modulating ionic fluxes in synapses. [1-2] Neurons are often considered to be the computational engines of the brain. [3] At the same time, synapses act far more than connection of neurons, and they are also responsible for massive parallelism, structural plasticity, fault-tolerant and robustness of the brain.[2, 3] The ability to simulate the functions of a brain will lead to a deeper understanding of neuroscience in a way that complements the physiological measurements of neural systems. Two approaches are available to implement neuromorphic simulation: software-based and hardware-based. Energy efficiency is a big challenge for the software-based approach since the algorithm is essentially run by conventional sequential machines with limited parallelism. For example, to perform a cortical simulation at the complexity of the human brain, the IBM's Blue Gene supercomputer demands heavy computation resources, requiring clusters of 1,572,864 microprocessors, 1.5 PB (1.5 million GB) of memories, and 6,291,456 threads, which consume an extremely high power of ~100 MW [4]. Alternatively, a hardware-based approach with massively parallel connections may overcome this challenge. Historically, synapses have been emulated by CMOS circuits with tens of transistors, which consume substantial area and energy and are not favorable for large-scale integration. [5]

Recently, mimicking synaptic and neural functions by nanoscale ionic/electronic devices with low power budgets has aroused world-wide interest. [6-12] Up to now, artificial synapses have been reported in $Ag_2S$ and $Cu_2S$ atom switching devices, Ag/a-Si and $WO_x$ memristors, $Ge_2Sb_2Te_5$-based phase change memories, hybrid organic and carbon nanotube transistors (CNT), etc. Synaptic plasticity, memory and some learning functions have been experimentally demonstrated. For example, Takeo Ohno et al successfully demonstrated the emulation of synaptic behaviors like short-term plasticity (STP) and long-term potentiation (LTP) in $Ag_2S$ atom switch. [6] Kyunghyun Kim et al reported a carbon nanotube (CNT) transistor recently. [12] The hydrogenation and dehydrogenation process in the CNT channel were found to be the reason for plasticity origin. Among all these artificial synapses, three-terminal devices are attractive because signal processing and learning in such devices can be performed simultaneously.[13]

Solid proton conducting electrolytes are attracting much attention due to their potential practical applications in clean energy fields such as fuel cells and electrolysis. Here, proton-conducting nanogranular phosphosilicate electrolytes gated oxide-based synaptic transistors with in-plane-gate figure [14, 15] were fabricated on glass substrates. Spike-timing-dependent plasticity (STDP), memory and learning functions of such artificial synapses were studied. Short-term memory (STM) to long-term memory (LTM) transition was demonstrated due to the electrochemical hydrogenation of the indium-zinc-oxide (IZO) channel. And most importantly, a simple artificial neural network with spatiotemporally correlated logic functions was mimicked in one synaptic transistor with dual in-plane-gate structure without any intentional hard-wire connections. The oxide-based artificial synaptic transistors proposed here could be the potential building blocks for artificial neural networks.

IZO-based synaptic transistors were fabricated on conducting ITO glass substrates at room temperature, as shown in Fig.1 (a). First, 900-nm-thick phosphosilicate films with nanogranular morphology were deposited on ITO glass substrate at room-temperature by plasma enhanced chemical-vapor deposition (PECVD) using $SiH_4/PH_3$ mixture and $O_2$ as reactive gases. Detailed structural and proton conductivity characterizations can be found in our previous publication [16]. Then, patterned IZO film arrays with the size of 150 μm ×1000 μm were deposited on the $SiO_2$-based electrolyte film by radio-frequency (RF) magnetron sputtering with a nickel shadow mask. IZO film deposition was performed using an IZO ceramic target ($In_2O_3$: ZnO =1:1 mol %) with a RF power of 100 W and a working pressure of 0.5 Pa. First, a 20-nm-thick IZO layer was sputtered in an $Ar/O_2$ mixed ambient with a flow rate of 14 sccm and 0.5 sccm for Ar and $O_2$, respectively. Then the oxygen gas was turned off, and a 30-nm-thick top IZO layer was deposited in ambient with gradually reduced oxygen. Electrical measurements of the synaptic devices were performed with a semiconductor parameter analyzer (Keithley 4200). All the measurements were done at room temperature with an air relative humidity (RH) of ~50%.

To test the synaptic response of the oxide-based artificial synapse, pre- and post-synaptic spikes were applied on the bottom ITO gate electrode and the patterned IZO channel film, respectively (Fig.1a). The post-synaptic currents were measured on patterned IZO with a constant bias of 0.5 V applied on two tungsten (W) probes. Typical temporal responses of the oxide-base artificial synapses to a pre-synaptic spike (1.0 V, 50ms) are shown in Fig. 1b. The pre-synaptic spike triggers an excitatory postsynaptic current (EPSC) above the resting current (~7.5 nA), reaching a peak value (~68 nA) at the end of the spike, and gradually reduces to the resting current. Such behaviors are similar to the EPSC in a biological excitatory synapse. When a positive pre-synaptic spike is applied on the gate ITO electrode (presynaptic neuron), protons in nanogranular $SiO_2$ electrolyte films will migrate and accumulate at the IZO/phosphosilicate electrolyte interface. At the end of the spike, the accumulated protons will diffuse back due to the concentration gradient. The migration of protons in the phosphosilicate film will induce an EPSC in the IZO channel due to the electrostatic couple effect.

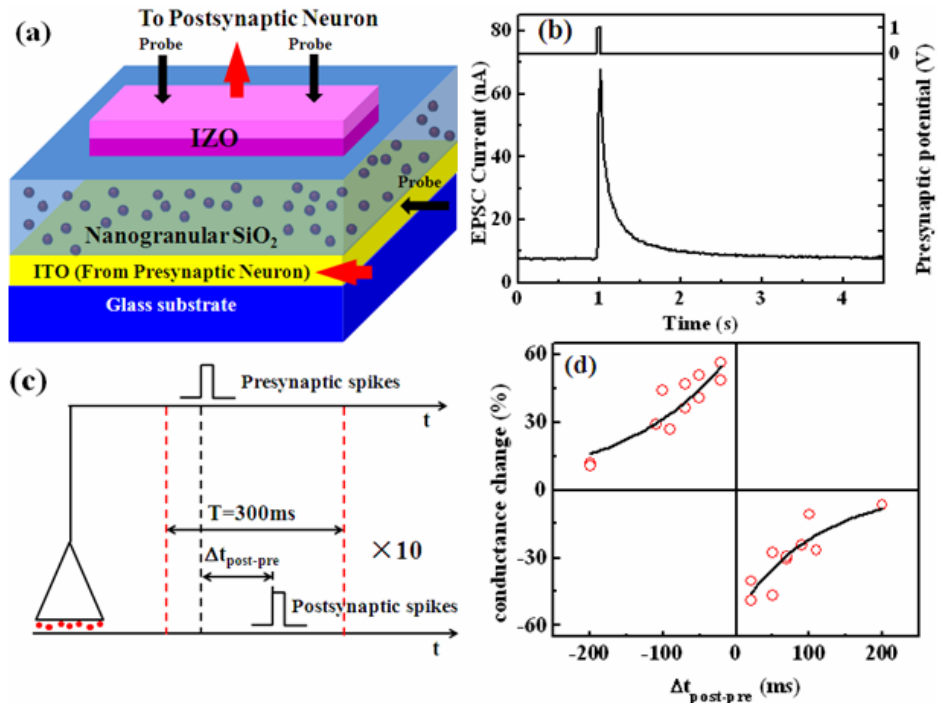

Figure 1 (a) The schematic image of the junctionless IZO-based synaptic transistor gated by nanogranular phosphosilicate electrolyte film. (b) The excitatory postsynaptic current (EPSC) plotted vesus time. A pre-synaptic spike was applied on the bottom ITO gate electrode, and the EPSC was measured by applying 0.5 V read voltage between two W probes connected to IZO channel. (c) 10 pairs of pre-synaptic spike and post-synaptic spike are repeated applied to the oxide-based synaptic transistor with a fixed inter-spike interval $\Delta t_{post-pre}$ for spike-timing dependent plasticity (STDP) stimulation. (d) STDP curves of the oxide-based synaptic transistor. The points are fitted with an exponential decay functions (black line).

In the biological system, spike-timing dependent plasticity (STDP) is essential to modify synapses in a neural network for learning and memory functions. [17-20] STDP is an elaboration of Hebbian learning applicable in spiking neural systems, in which the modulation of synaptic weight is based on the relative timing of spikes produced by the pre-synaptic and postsynaptic neurons. Following the test protocol in neurological experiments, 10 pairs of pre-synaptic spike (4.0 V, 20ms) and post-synaptic spike (5.0 V, 20ms) with a fixed inter-spike interval, $\Delta t_{post-pre}$, were applied on the oxide-based synaptic transistor (Fig. 1c). During the test, conductance of the IZO channel was defined as synaptic weight. Channel conductance was measured by applying read pulse (0.2 V, 50ms) between two W probes connected with the IZO channel before ($G_0$) and 1 min after ($G_t$) the spike-pair application. The relative change of synaptic weight is defined as:

$$\Delta W = (G_t - G_0)/G_0$$

The changes in synaptic weight ($\Delta W$) were measured under different $\Delta t_{post-pre}$ ranging from -200 ms to 200 ms as shown in Fig. 1d. The data in Fig. 1d are statistically scattered, similar to that observed in biological synapses. [19] The changes in the synaptic weigh ($\Delta W$) could be fitted by an exponential function $\Delta W = A_+ \exp(\Delta t_{post-pre}/\tau_+)$ for $\Delta t_{post-pre} < 0$ and $\Delta W = A_- \exp(\Delta t_{post-pre}\backslash\tau_-)$ for $\Delta t_{post-pre} > 0$. $A_+$ and $\tau_+$ are fitted to be ~ 63% and 145 ms, respectively. And $A_-$ and $\tau_-$ are fitted to be ~-56% and 106 ms, respectively. The A and τ values are comparable with those observed in biological synapses. [17-19]

In neocortical slices, hippocampal slice and cell cultures, long-term weakening of synapses occurs when pre-synaptic action potentials follow post-synaptic firing. While pre-synaptic action potentials that precede post-synaptic spikes produce long-term strengthening of synapses. The largest changes in synaptic efficacy occur when the time difference between pre- and post-synaptic action potentials is small. There is a sharp transition from strengthening to weakening as this time difference passes through zero. As reported previous, phosphosilicate electrolyte with nanogranular structure is an inorganic proton conductor. [21-22] When a high positive gate voltage is applied on ITO gate electrode, protons in the phosphosilicate electrolyte can penetrate into the IZO channel, which will result in a permanent increment in channel conductance due to electrochemical reaction. [23, 24] When the post-synaptic spike is applied shortly after the pre-synaptic spike ($\Delta t_{post-pre}>0$), protons penetrated into the IZO channel can be driven back into phosphosilicate electrolyte, and protons in IZO film is depleted after several spike pairs. Eventually, a permanent reduction of IZO conductance can be observed, which is similar to the long term depression (LTD) in vivo. On the contrary, when the pre-synaptic spike is applied shortly after the post-synaptic spike ($\Delta t_{post-pre}<0$), protons in the phosphosilicate electrolyte will penetrate into the IZO channel, and more and more protons is accumulated in the IZO film after several spike pairs. At last, a permanent increase of IZO conductance can be observed, which is similar to the long-term potentiation (LTP) in vivo. In both cases, spike pairs with a shorter $|\Delta t_{post-pre}|$ can induced a more remarkable change in IZO channel conductance (or synaptic weight).

To test the effects of the repeated pre-synaptic spikes on the synaptic weight, pre-synaptic spikes (2.0 V, 50 ms) with a pulse-to-pulse interval of 50 ms were applied on ITO gate electrode. Different spike numbers (N=1, 2, 5, 10, 20, 50) were applied, starting from the same initial state (0.32 μS). Figure 2a shows the changes of the IZO channel current obtained by applying 5 input pre-synaptic spikes (2.0 V, 50 ms) and a constant read voltage of 0.5 V between two W probes on IZO. The channel current progressively increases with the pulse number and decays back to a static equilibrium value in a short time. Fig. 2b shows the retention time ($\tau$) plotted with respect to the number of stimulations (N). The retention time is defined as the time from removing the spikes to the channel conductance decay back to the original value. An obvious increase in $\tau$ is observed (the red line) with repetitive stimulations. The residual conductance ($G_R$) of IZO film versus pulse number (N) is also plotted in Fig. 2b (the black line). The residual conductance is defined as the conductance immediately after the spikes. Short-term memory (STM) is the capacity for holding a small amount of information in mind in an active, readily available state for a short period of time (generally in the order of seconds). Therefore, our results could be analogous to the short-term memory (STM). The conductance can have an increment for few seconds after gate pulse, which corresponds to the temporal enhancement in memory retention after repetitive stimulations.

At the same time, the conductance decay of IZO channel suggests a close similarity to "the forgetting curve" proposed by Ebbinghaus in 1885 [25]. Such curves demonstrate that the repetition rehearsal is an appropriate method for increasing memory strength. Figure 2c shows the normalized conductance decay curves. A power function($y=b\times t^{-m}$), generally used to analyze psychological behavior such as STM [26], was used to fit the conductance curves. In the power function, y is the memory retention, b is the fit constant for scaling, t is the time from the end of the rehearsal and m is the power function rate. Memory retention was normalized using the residual conductance value ($G_R$). The factor m is observed to decrease with the increase in the number of rehearsals, indicating that the forgetting rate is slowed down by active recalls.

In order to mimic the long term memory process in the oxide-based synapses. Fifty pre-synaptic spikes (8.0 V, 50 ms) were applied on ITO gate electrode. The synaptic weight is measured by applying a constant voltage of 0.5 V between two W probes connected to IZO channel film. As shown in Fig. 2d, the channel current dramatically increases with the increment of input pulse number and slowly decayed back to a static equilibrium value which is much higher than the original channel current value. In this case, both the residual channel conductance and the retention time are quite larger than the STM mode. The residual channel conductance is three orders of magnitudes higher than the initial value, and the retention time ($\tau$) can exceed one hundred year. In biological systems, long term memory (LTM) can last from hours to days or even to a life time.[27] The acquisition of the LTM is dependent upon the construction of new proteins, through intricate process involving many molecular mechanisms and structural changes at various cellular levels.[28, 29] Such LTM process include particular transmitters, receptors, and new synapse pathways that reinforce the communicative strength between neurons. In our device, the persistent high conducting state of IZO layer is due to the proton-related electrochemical process under high gate pulses.

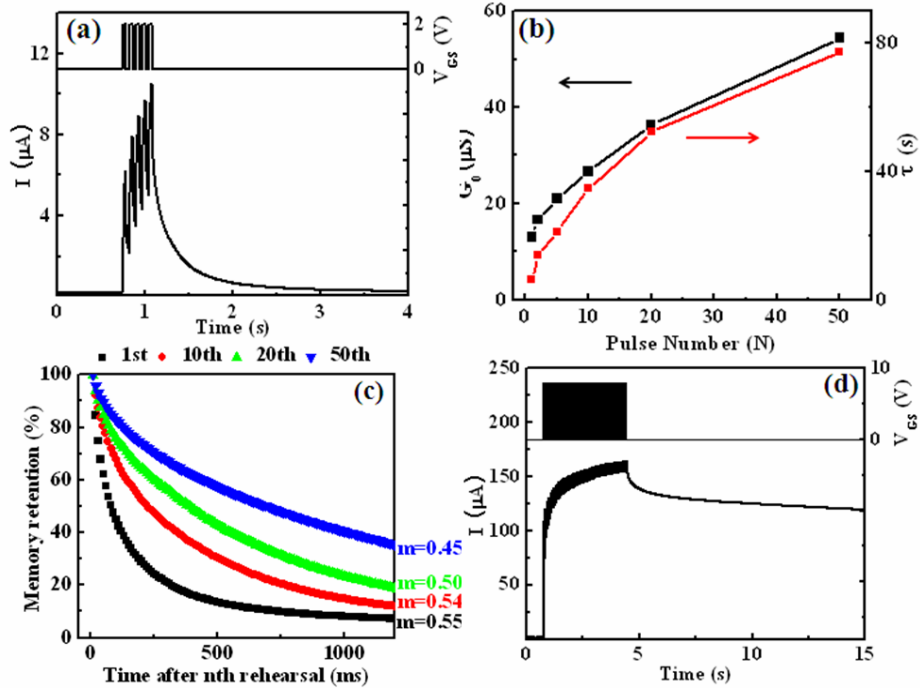

Figure 2. (a) The changes in current obtained by applying 5 input pulses (2.0 V, 50 ms) on ITO gate electrode and a constant voltage of 0.5 V between two W probes connected to IZO channel (b) The residual conductance ($G_R$) of IZO channel and retention time ($\tau$) plotted with respect to the number of stimulations (N). (c) Normalized conductance decay curves. (d) The changes in current obtained by applying 50 input pulses (8.0 V, 50 ms) on ITO film with 0.5 V bias applied on IZO channel film. The maximum current can up to 164 μA and the retention time can exceed 100 years.

When a presynaptic neuron receives two stimuli in rapid succession, the postsynaptic response will commonly be larger for the second than for the first pulse — a phenomenon known as paired-pulse facilitation (PPF). [30] As a common form of short-term synaptic plasticity in biological synapses, paired pulse facilitation (PPF), reflect a responsivity for the second pulse when first pulse and the second pulse are close enough.[31] Such PPF is essential to decode temporal information in auditory or visual signals. In our case, oxide-based synapse can process temporally correlated spikes and generate temporal analog logic such as PPF. Two successive pre-synaptic spikes (1.0 V, 50 ms) were applied on pre-synaptic neuron (gate electrode). The inter-spike interval ($\Delta t_{pre}$) ranges from 50 ms to 5000 ms. Fig.3 (a) shows the EPSC triggered by a pair of pre-synaptic spikes with $\Delta t_{pre}$=250 ms. The EPSC triggered by the second pre-synaptic spike is larger than the that triggered by the first spike. The ratio of the amplitudes between the second EPSC (A2) and the first EPSC (A1) is plotted versus $\Delta t_{pre}$ in Figure 3 (b). The ratio gradually decreases with increasing $\Delta t_{pre}$. A maximum value of 137% is obtained at the $\Delta t_{pre}$=50 ms. The inset of Fig. 3(b) illustrates the schematic diagram for PPF measurement. After the first spike, the protons in phosphosilicate electrolyte would drift back to their equilibrium positions due to the concentration gradient. If the second spike is applied close enough, the protons triggered by the first spike still partially reside near IZO channel. Thus, the protons triggered by the second spike near IZO are augmented with the residual protons. Bigger $\Delta t_{pre}$ would induce less residual hydrogen ions near IZO channel and result in lower value of A2/A1.

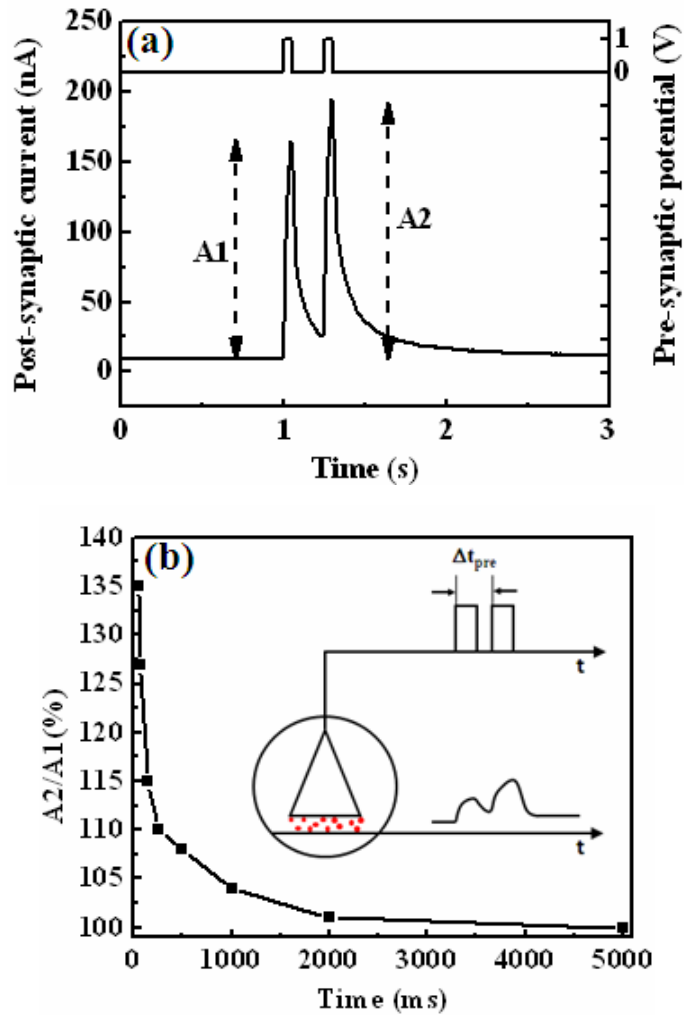

Figure 3. (a) The EPSC triggered by a pair of pre-synaptic spikes with $\Delta t_{pre}$=250 ms. The EPSC triggered by the second pre-synaptic spike is larger than the EPSC by the first spike. A1 and A2 was define as the amplitudes of the first and second EPSCs, respectively. (b) The ratio of the amplitudes (A2/A1) between the second EPSC and the first EPSC is plotted versus $\Delta t_{pre}$. The ratio decreases with the inreasing interval-time between the two spikes. The inset is the Schematic diagram of the measurment of paired-pulse facilation (PPF). Two successive pre-synaptic spikes (1 V, 50 ms) were applied on pre-synaptic neuron. The inter-spike interval ($\Delta t_{pre}$) ranges between 50 ms and 5000 ms.

The highly parallel process in the neuron networks is mediated through a mass of synaptic interconnections. Pre-synaptic spikes from different neurons can trigger a post-synaptic current through synapses in a post-synaptic neuron to establish dynamic logic in a neural network. [32, 33] Spatiotemporally correlated dynamic analog logic can also be mimicked in our dual in-plane-gate oxide-based protonic/electronic hybrid synaptic transistors. As shown in Fig. 4 (a), the left two patterned IZO films (T1 and T2) can be regarded as the terminals of pre-synaptic neurons (neuron 1 and neuron 2) and the right patterned IZO film (T3) is defined as the terminals of post-synaptic neuron. Therefore, a simple artificial neural network is obtained with two synapses. Fig. 4 (b) shows the schematic diagram for measuring the spatiotemporally correlated function. Two isolated pre-synaptic spikes (0.5 V, 50 ms) with an inter-spike interval ($\Delta t_{pre2-pre1}$) come from two different pre-synaptic neurons, respectively. The EPSCs (EPSC1 and EPSC2) triggered by such two spikes are summed in the post-synaptic neuron (neuron 3). Fig. 4 (c) shows the EPSCs triggered by the spatiotemporally correlated two spikes at $\Delta t_{pre2-pre1}$=-500 ms, 0 ms, 500 ms. When $\Delta t_{pre2-pre1}$= 0, two spikes from different neurons are applied on pre-synaptic neurons simultaneously. The amplitude of the EPSC in the post-synaptic neuron reaches the maximum value. When $|\Delta t_{pre2-pre1}|$= 500 ms, the two EPSCs seems have no detectable influence on each other. As shown in Fig. 4 (d), the amplitude of EPSC at t = 0 (as soon as the pre-synaptic spike applied on T1 ended) decreases asymmetrically with increasing $|\Delta t_{pre2-pre1}|$. When EPSC1 from the neuron 1 is triggered earlier than EPSC2 from the neuron 2 ($\Delta t_{pre2-pre1}$> 0), the amplitude of EPSC at t = 0 is equal to the peak amplitude of EPSC1. When EPSC1 from the neuron 1 is triggered later than EPSC2 from the neuron 2 ($\Delta t_{pre2-pre1}$<0), the amplitude of EPSC at t = 0 is the summed current of EPSC1 from neuron 1 and EPSC2 from neuron 2. When $\Delta t_{pre2-pre1}$ decreases further ($\Delta t_{pre2-pre1}<0$), the influence of EPSC2 from synapse 2 is gradually less significant.

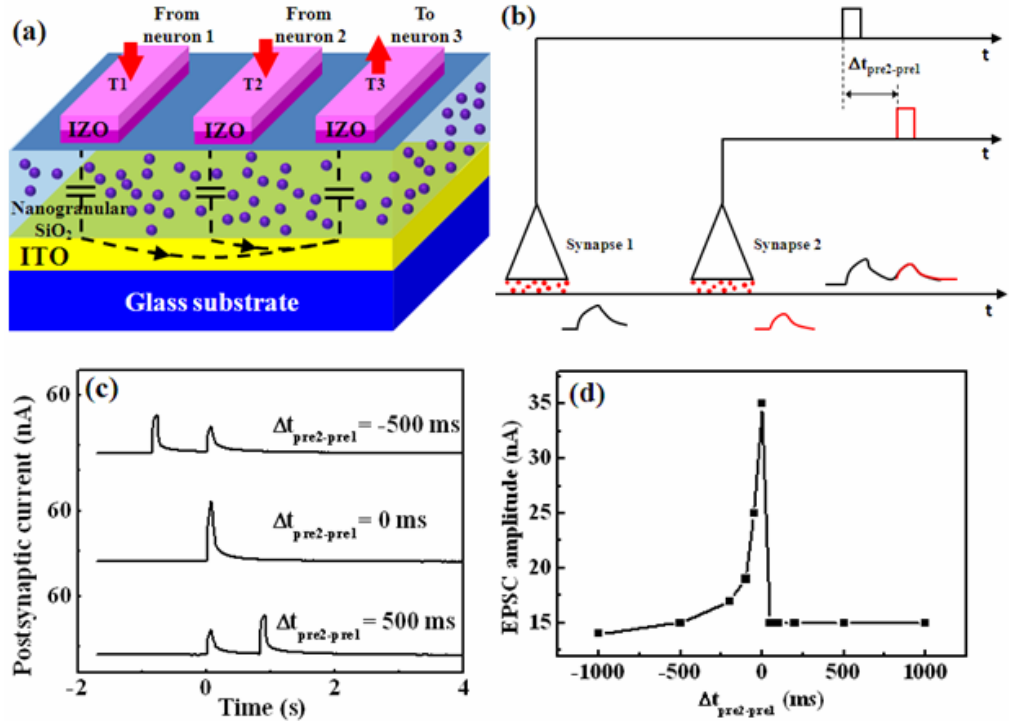

Figure 4 (a) The structure of the simple artificial network with three neurons. (b) The schematic diagram for measuring the spatiotemporally correlated function. The two pre-synaptic spikes (1 V, 50 ms) with an inter-spike interval ($\Delta t_{pre2-pre1}$) ranging from -1000ms to 1000ms, were applied on T1 and T2, respectively, and EPSCs was measured on T3. (c) EPSCs plotted versus time with two spatiotemporally correlated spikes at $\Delta t_{pre2-pre1}$=-500ms, 0ms, 500ms. (d) The amplitude of the EPSC at t=0 (as soon as the pre-synaptic spike applied on T1 ended) is plotted as a function of $\Delta t_{pre2-pre1}$ between the two pre-synaptic spikes.

**Conclusions**

In summary, proton-conducting phosphosilicate electrolytes gated oxide-based protonic/electronic hybrid artificial synapses were fabricated on glass substrates at room temperature. Synaptic dynamic functions such as excitatory post-synaptic current triggered by pre-synaptic spike and spike-timing dependent plasticity were mimicked in such artificial synapse. Short-term memory (STM) to long-term memory (LTM) transition was realized by tuning pre-synaptic spike voltage amplitude, and LTM was due to the proton-related interfacial electrochemical reaction. Most importantly, spatiotemporally correlated functions were realized on a simple neural network with integration of three neurons and without any intentional hard-wire connections. The amorphous oxide-based artificial synapses coupled by proton-conducting phosphosilicate electrolytes demonstrated here could be used as the potential building blocks for artificial neural networks.